\documentclass[structabstract]{aa}  
\usepackage{graphicx}
\usepackage[authoryear]{natbib}
\usepackage{amsmath}
\usepackage{amssymb}
\usepackage{txfonts}
\usepackage{threeparttable}
\begin{document}
   \title{Pulse-amplitude-resolved spectroscopy of bright accreting pulsars: 
     indication of two accretion regimes}


   \author{D. Klochkov\inst{1}\and
     R. Staubert\inst{1}\and
     A. Santangelo\inst{1}\and
     R. E. Rothschild\inst{2}\and
     C. Ferrigno\inst{3}
         }

   \institute{Institut f\"ur Astronomie und Astrophysik, Universit\"at 
     T\"ubingen (IAAT), Sand 1, 72076 T\"ubingen, Germany
     \and
     Center for Astrophysics and Space Sciences, University of California, San Diego, La Jolla, CA, USA
     \and
     ISDC Data Center for Astrophysics of the University of Geneva	 
     chemin d'\'Ecogia, 16 1290 Versoix, Switzerland
   }

   \date{Received ***; accepted ***}

  \abstract
   {In addition to coherent pulsation, many accreting neutron stars exhibit
    flaring activity and strong aperiodic variability on time scales
    comparable to or shorter than their pulsation period. Such a behavior
    shows that the accretion flow in the vicinity of the accretor must be
    highly non-stationary. Observational study of this phenomenon is often 
    problematic as it requires very high statistics of X-ray data and a 
    specific analysis technique.}
   {In our research we used high-resolution
     data taken with \textsl{RXTE} and \textsl{INTEGRAL}
     on a sample of bright transient and persistent
     pulsars, to perform an in-depth study of their variability on time scales
     comparable to the pulsation period -- ''pulse-to-pulse
     variability''.}
   {The high-quality data allowed us to collect individual
     pulses of different amplitude and explore their X-ray spectrum 
     as a function of pulse amplitude.
     The described approach allowed us for the first time to study 
     the luminosity-dependence of pulsars' X-ray spectra in observations where 
     the averaged (over many pulse cycles) luminosity of the source remains 
     constant.}
   {In all studied pulsars we revealed significant spectral changes
     as a function of the pulse height both in the continuum 
     and in the cyclotron absorption features. 
     The sources appear to form two groups showing 
     different dependencies of the spectrum on pulse height. 
     We interpret such a division as a manifestation of two distinct
     accretion regimes that are at work in different pulsars.}
   {}


   \maketitle
%

\section{Introduction}

Strong aperiodic variability of the X-ray flux is a well known
characteristic of many X-ray binary systems. First discovered in
black hole candidates (with Cyg~X-1 being the most remarkable case),
it was later shown to be a common feature also among accreting neutron stars
\citep[see e.g.][for a review]{Belloni:Hasinger:90}. Although \emph{homogeneity}
and \emph{stationarity} of the accretion flow is often assumed in calculations
dealing with physics inside the X-ray emitting structure on a neutron star
(as it greatly simplifies the mathematical treatment of the problem),
the observed variability clearly indicates that the accretion flow 
in the close vicinity of accretor must be highly non-stationary.
On the other hand, the 
penetration of matter into the magnetosphere of the neutron
star around the stopping radius is believed to be governed by various
kinds of instabilities which plasma supported by magnetic field is subject to
\citep[see e.g.][]{Ghosh:Lamb:79,Kulkarni:Romanova:08,Bozzo:etal:08}.
The instabilities will naturally lead to fragmentation of a continuous 
flow into more or less isolated blobs or filaments.
More focused theoretical studies of the non-stationary accretion problem
have been performed e.g. by \citet{Morfill:etal:84,Demmel:etal:90,
Orlandini:Boldt:93} who have explicitly shown that the inhomogeneity
of the flow is generally expected in accreting pulsars and that it does not
only arise from the original inhomogeneity of the accreted matter
(e.g. ''clumps'' in the stellar wind or variations of matter supply from
the donor star) but can naturally be
produced by instabilities close to the
magnetospheric boundary of the neutron star.

So far, the aperiodic variability of accreting pulsars has mostly been
studied by means of power spectra of their high time-resolution light curves
\citep[e.g.][]{Belloni:Hasinger:90,Pottschmidt:etal:98,Revnivtsev:etal:09}. 
In our work, however, we concentrate on the variability of 
individual X-ray pulses
often referred to as \emph{pulse-to-pulse variability}.
Such kind of variability seems to be a common phenomenon among 
accreting pulsars and has been reported by several authors for 
bright outbursts of some transient or strongly variable
sources: \citet{Frontera:etal:85} for A\,0535+26, 
\citet{Staubert:etal:80,Kretschmar:etal:00}
for Vela\,X-1, \citet{Tsygankov:etal:07} for 4U\,0115+63. However, a
detailed study of the pulse-to-pulse variability is usually limited by
the photon statistics, especially for relatively fast pulsars (with a
spin period of a few seconds or less). 
In the present work, we used the publicly available
archival data from high-sensitivity X-ray detectors onboard
\textsl{RXTE} and \textsl{INTEGRAL} taken on a set of 
four bright accreting pulsars --
V\,0332+53, 4U\,0115+63, A\,0535+26, and Her\,X-1. 

The four accreting pulsars are well established
cyclotron line sources, i.e. their spectra contain
\emph{Cyclotron Resonant Scattering Features (CRSFs)}.
Such features appear as absorption lines due to resonant scattering 
of photons off the relativistic plasma electrons at Landau levels 
\citep[see e.g.][]{Truemper:etal:78,Isenberg:etal:98,Araya:Harding:00}.
The CRSFs, if detected, provide a direct way to measure the magnetic
field strength at the site of X-ray emission as the energy of the
fundamental line and the spacing between the harmonics are proportional
to the $B$-field strength.
In some sources the cyclotron line energy was found to change with
luminosity. In V\,0332+53 and 4U\,0115+63 a negative correlation
of the line energy with luminosity was observed
\citep{Tsygankov:etal:07,Tsygankov:etal:10,Mowlavi:etal:06,Mihara:etal:04}
while for Her~X-1, \citet{Staubert:etal:07} reported a positive correlation
of the line energy with luminosity.

A description of the
observational data is provided in Sect.\,\ref{sec:obs}. Using a special analysis
technique (described in Sect.\,\ref{sec:techn}), we collected individual 
pulses of different amplitudes and studied the 
differences in their X-ray spectra.
As a result, we revealed significant correlations of different spectral
parameters with pulse amplitude (Sect.\,\ref{sec:results}).
Based on the sign of the correlations the explored pulsars
can be divided into two groups. 
We argue that the two groups correspond to
two distinct regimes of
accretion that are at work in different sources:
local sub- and super-Eddington regime (see discussion in
Sect.\,\ref{sec:discussion}). The summary and conclusions are
provided in Sect.\,\ref{sec:summary}.

\section{Observations}
\label{sec:obs}

For our analysis we used the data taken with the \textit{RXTE} satellite
\citep{Bradt:etal:93}
during the intense outbursts of the transient high mass X-ray binaries (HMXB)
V\,0332+53, 4U\,0115+63, and A\,0535+26 and during a \emph{main-on} state of the
persistent intermediate mass X-ray binary Her~X-1 (such states of
high X-ray flux repeat in the system about every 35\,days most probably
reflecting periodic obscuration of the source by a precessing tilted
accretion disk, see e.g. \citealt{Klochkov:etal:06} and references
therein). In the case of A\,0535+26 we also used data taken with the
\textsl{INTEGRAL} observatory \citep{Winkler:etal:03}
simultaneously with the \textsl{RXTE} observations.
The main X-ray instruments onboard the two satellites
provide a broad-band coverage of the sources' X-ray spectra:
\textit{RXTE/PCA} \citep{Jahoda:etal:96} and \textit{INTEGRAL/JEM-X}
\citep{Lund:etal:03} are sensitive from a few keV to $\sim$35\,keV, while
\textit{RXTE/HEXTE} \citep{Rothschild:etal:98} and
\textit{INTEGRAL/IBIS(ISGRI)} \citep{Ubertini:etal:03} --
from $\sim$20\,keV to a few hundred keV.

Our main criterion for the selection of the observations 
and instruments was a sufficiently dense timing
coverage of the sources during their high flux states which provides
best statistics. Additionally, for \textit{RXTE/PCA} we searched for the data
taken in an appropriate data mode which allows one to achieve 
simultaneously high time- and
energy-resolution in the entire \textit{PCA} energy range. 
Specifically, we used the data taken in 
\textit{GoodXenon} and certain \textit{Generic Event} modes
(with sufficient energy- and time-resolution). 
Where possible, we selected the \textit{RXTE} observations with both
\textit{HEXTE} clusters switched on to maximize the photon statistics.
\textit{INTEGRAL} data were only used in case of A\,0535+26 which
has a long ($\sim$100\,s) pulse period. For other three pulsars 
with much shorter pulse periods (a few seconds), individual pulsations 
cannot be distinguished in the \textit{INTEGRAL (IBIS/ISGRI)} light curves,
thus, preventing pulse-to-pulse study.

\begin{table}
  \centering
  \caption{Observations used for the pulse-amplitude-resolved
    analysis}
  \label{obs}
  \begin{threeparttable}
    \vspace*{1mm}
    \begin{tabular}{l l l l}
      \hline\hline
      Source name & Instrument   & mid MJD & Exposure (ksec) \\
      \hline
                  &                   &                &      \\
      V\,0332+53  & \textsl{RXTE}     & 53354\tnote{a} & 23.7 \\
      4U\,0115+63 & \textsl{RXTE}     & 51249\tnote{b} & 32.8 \\
      A\,0535+26  & \textsl{RXTE}     & 53615\tnote{c} & 30.8 \\
                  & \textsl{INTEGRAL} & 53615\tnote{c} & 104.7\\
      Her X-1     & \textsl{RXTE}     & 52600\tnote{d} & 98.7 \\
      \hline
    \end{tabular}
    \begin{tablenotes}
    \footnotesize \item [a] Giant outburst in 2004
    \item [b] Giant outburst in 1999
    \item [c] Normal outburst in 2005
    \item [d] Main-on state
    \end{tablenotes}
  \end{threeparttable}
\end{table}

For each of the three transient sources we used a short continuous set 
of pointings covering $\sim$1 day or less of the brightest part of an
outburst (close to its maximum) so that the average flux level 
within the observations did not change significantly. The observed
flux variability is therefore related to X-ray pulsations and
pulse-to-pulse variations.
For the persistent pulsar Her~X-1 we used the data from the
main-on state of the source corresponding to the 35\,d cycle No.\,323 
(according to the numbering convention adopted in \citealt{Staubert:etal:83})
which is best covered by \textsl{RXTE} observations.
We selected the data from the middle part of the main-on
where the flux does not change significantly, i.e. where the obscuration
by the accretion flow is minimal. 
The chosen observations of Her~X-1 are spread over 
$\sim$5\,d. A summary of the observational data 
on all the sources is provided in Table\,\ref{obs}

\section{Analysis technique\label{sec:technique}}
\label{sec:techn}

The standard data reduction has been performed with
the software packages and calibration data provided by the 
instrument teams. For \textit{RXTE} we used 
HEASoft\,6.9\footnote{\texttt{http://heasarc.nasa.gov/lheasoft/}},
for \textit{INTEGRAL} -- 
OSA\,9.0\footnote{\texttt{http://www.isdc.unige.ch/integral/analysis\#Software}}.

\begin{figure}
\resizebox{\hsize}{!}{\includegraphics{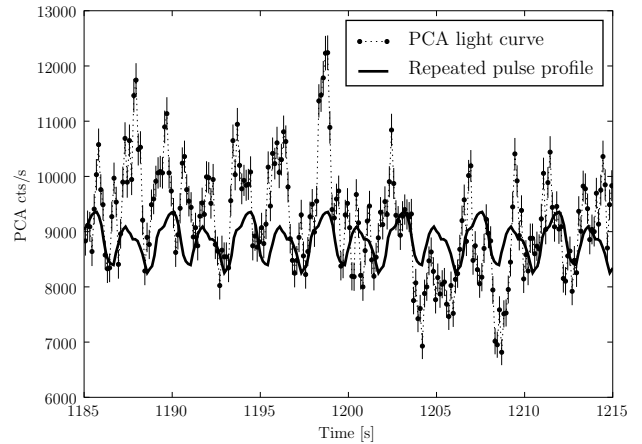}}
\caption{Sample light curve of V\,0332+53 during its 2005 giant outburst 
  obtained with \textsl{RXTE/PCA} and summed over all energy channels
  (2--80\,keV). 
  The solid curve shows the repeated pulse profile obtained by folding
  a longer data sample. Dramatic pulse-to-pulse variability is clearly seen.
}
\label{lc_p2p}
\end{figure}

\begin{figure}
\resizebox{\hsize}{!}{\includegraphics{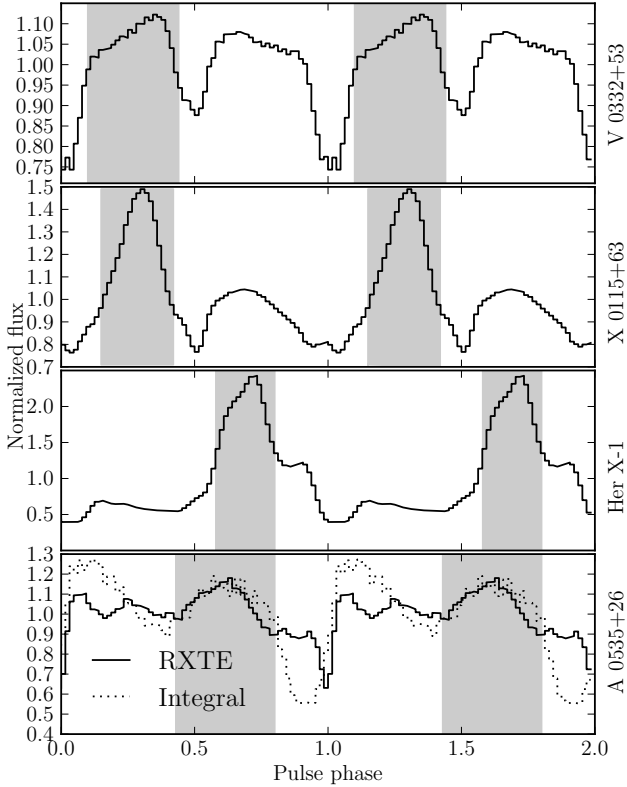}}
\caption{Averaged (over many pulsation cycles) pulse profiles of the 
pulsars in our sample obtained with \textit{RXTE/PCA} in 2--80\,keV range 
(solid curve) and (in case of A\,0535+26) with 
\textit{INTEGRAL/ISGRI} in 20--100\,keV range (dotted curve on 
the bottom panel).
The shaded areas mark the pulse-phase intervals used for our
pulse-amplitude-resolved analysis (see text).}
\label{pp}
\end{figure}

To analyze the pulse-to-pulse variability, for each source we extracted
a high-resolution light curve where single pulses are clearly
distinguishable. An example of such a light curve obtained with
\textsl{RXTE/PCA} in 2--80\,keV range 
(this is the entire \textit{PCA} range, although the effective area
of the instrument drops above 30--35\, keV)
on V\,0332+53 is shown in Fig.\,\ref{lc_p2p} where
the strong variability of the profile shape from one pulse to the next
is visible. We then selected a pulse-phase interval
containing the brightest part of the pulse profile,
referred to as \emph{pulse}. 
The chosen pulse-phase intervals for each source
are shown in Fig.\,\ref{pp} by the shaded areas.
For each pulse we calculated its mean count rate
(in all \textit{PCA} channels for the \textit{RXTE} data and
in 20--100\,keV for the \textit{INTEGRAL} data), that is
the average count rate within the selected pulse-phase
interval, which we call the \emph{amplitude} or \emph{height}
of the pulse.
As can be seen in Fig.\,\ref{lc_p2p} the amplitude of individual pulses
varies over a broad range. This allows one to explore the variation of
the X-ray spectrum as a function of pulse amplitude, which is the
central point of our research. 

The photon statistics does not allow us to extract meaningful spectra
of single pulses. Therefore, we grouped together pulses of similar
amplitude. For this purpose we explored the distribution
of pulse amplitudes for each source and split the entire range
of amplitudes into five to six bins, keeping approximately the 
same statistics within each bin. Then, for each amplitude bin
we constructed a list of \emph{Good Time Intervals} (GTIs) to select the
pulses (i.e. the data inside the selected pulse-phase intervals)
whose amplitude falls into the bin. 
Providing the data-processing pipeline with the produced GTIs 
we extracted the broad band X-ray spectra for the selected bins, i.e. as a 
function of pulse height. 
We note again that the extracted spectra correspond
to a fixed pulse-phase interval which minimizes possible influence
of pulse-phase dependence of the spectrum.

To maximize the photon statistics in our spectra, where possible 
we used \textit{PCA} data up to 60\,keV. Even though the effective
area of the instrument 
at this energy drops dramatically along with an increase of 
the relative background contribution, the photon statistics is still
competitive with that of \textit{HEXTE} (which suffers from larger
dead time). 
For the spectral analysis of the \textit{PCA} data we used the updated
response v11.7 (2009 May 11) and background estimation
files \textit{Sky\_VLE}. According to the instrument team, the new 
calibration files in combination with the updated \textit{HEASoft} 
package (starting from version 6.7) provide significant improvements 
to spectral analysis and allow to use \textit{PCA} data up to
$\sim$50\,keV with systematic uncertainties of 0.5\%\footnote{http://www.universe.nasa.gov/xrays/programs/rxte/pca/doc/rmf/pcarmf-11.7/}. 
Our analysis of pulse-averaged spectra between 40 and 60\,keV
has shown that the \textit{PCA} spectrum extracted with the new
calibration is in complete accord with the \textit{HEXTE} spectrum
for which the background is directly measured during the observations
(using the off-set pointings of the collimator). The agreement
between the two detectors is also claimed by \citet{Rothschild:etal:11}
who also used \textit{PCA} data up to 60\,keV analyzing the RXTE observations
of Cen\,A. Following \citet{Rothschild:etal:11}, the background model counts
histogram was included in the spectral fitting as a correction file
in order to account for small deviations of the background
and dead-time models from reality (\texttt{recorn} model in
\textit{XSPEC\,12.6.0}). We added systematic uncertainties 
at the level of 1.0\% to our \textit{PCA} spectra which are somewhat 
larger than the recommended 0.5\% (see above) but allow us to 
obtain more conservative estimates of the spectral parameter errors.
No systematic errors have been added to \textit{HEXTE} as it is not 
recommended by the instrument team (the uncertainties of this instrument 
are dominated by statistical fluctuations).

For the \textit{INTEGRAL} instruments we used the standard method
of background evaluation from the deconvolution of the detector images
provided by the \textit{OSA} software\footnote{http://www.isdc.unige.ch/integral/download/osa\_doc}. We added systematic uncertainties 
at the level of 2\% to the \textit{JEM-X} spectra and 1\% to the
\textit{IBIS/ISGRI} spectra based on the recommendations of the 
instrument teams and the Crab observations closest to our
observations.

It is known that inaccuracy in the absolute calibration of the X-ray 
instruments onboard 
\textit{INTEGRAL} and \textit{RXTE} might lead to systematic differences
in the observed spectral shape of a source between the two satellites
\citep[see e.g.][]{Tsujimoto:etal:11}. In our research, however, we focus
on \emph{relative} variations of the spectral parameters as a function of 
pulse amplitude rather than their absolute values. Therefore, the
cross-calibration accuracy is not critical for our study. 
We also note that the pulse-averaged cyclotron line energies measured 
simultaneously with \textit{INTEGRAL} and \textit{RXTE} in A\,0535+26
\citep{Caballero:etal:07} and Her\,X-1 
\citep{Klochkov:etal:08b,Staubert:etal:07} are in agreement within 
$\sim$1\,keV between the two satellites.

\section{Results}
\label{sec:results}

For each pulsar in our sample we obtained a series of broad band
($\sim$3--80\,keV) X-ray spectra corresponding to different
pulse amplitudes. The spectral continua are modeled using the
powerlaw-cutoff function. The fluorescent K$_\alpha$ emission at
6.4\,keV has been modeled by an additive Gaussian. 
The absorption cyclotron features are clearly seen in our
pulse-amplitude-resolved spectra and are
modeled using multiplicative absorption lines with a Gaussian
optical depth profile. 
More detailed description of the spectral models
is provided below, in the subsections devoted to the individual sources.
As the analysis procedures are very similar for all
sources in our sample, we provide a full description of our spectral fitting
(including a list of spectral parameters, $\chi^2$ values etc.)
only for the first pulsar, V\,0332+53 (Sect.\,\ref{sec:v0332}).
For the other the sources we skip the detailed information 
and concentrate on final results -- variations of the spectral
parameters with pulse amplitude.

All our spectral fits were checked for possible intrinsic
dependencies/degeneracies of different parameter pairs using error contour 
plots. No significant correlations were found except the one between 
the photon index and the cutoff energy which was eliminated by fixing
the cutoff energy (see below).
The presented
results are also found to be stable with respect to the choice of
different spectral functions, both for the continuum and the cyclotron
absorption lines. We are, therefore, confident
that the variability reported below arises from the sources behavior 
and reflects real physics.

\subsection{V\,0332+53\label{sec:v0332}}
\label{subsec:v0332}

V\,0332+53 belongs to the category of Be/X-ray binaries (BeXRB) -- currently 
the most numerous class of HMXBs. The neutron star in such systems
periodically or sporadically accretes mass from the equatorial disc
around the Be- or Oe-type optical companion. The episodes of accretion
give rise to powerful X-ray outbursts while in quiescence the X-ray flux
often falls below the detection limit of most instruments.
The neutron star in V\,0332+53 has a spin period of $\sim$4.3\,s
\citep{Stella:etal:85}.
For this source we used the \textit{RXTE} data close to the 
maximum of its giant (type II) outburst in 2004 
\citep[see e.g.][for the nomenclature of outburst types in BeXRBs]{Coe:00},
when the bolometric luminosity reached a few times $10^{38}$\,erg\,s$^{-1}$
for a distance of 7\,kpc \citep{Tsygankov:etal:10}.
Up to three cyclotron features have been detected in the source: 
the fundamental line at $\sim$26\,keV and two harmonics, at 
$\sim$50 and $\sim$70\,keV respectively. 
As mentioned in the Introduction,
during the outbursts, the fundamental 
line exhibits a strong negative correlation with the X-ray flux
\citep[e.g.][]{Tsygankov:etal:10,Mowlavi:etal:06}.

\begin{figure*}
\centering
\includegraphics[width=17cm]{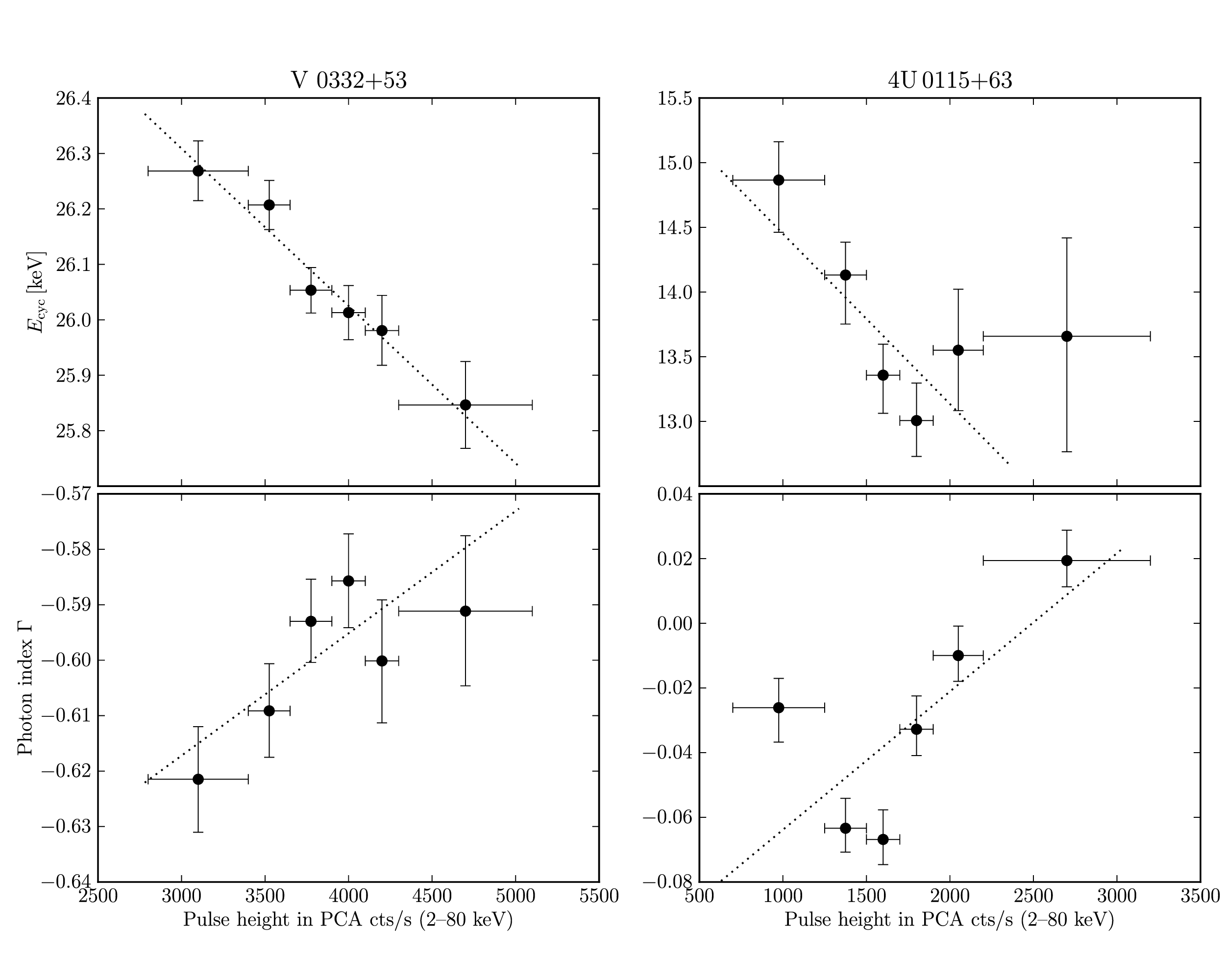}
\caption{The variation of the fundamental cyclotron line centroid energy 
$E_{\rm cyc}$ (top) and the photon index $\Gamma$ (bottom)
with pulse amplitude measured in V\,0332+53 (left)
and in 4U\,0115+63 (right) using the 
\textsl{RXTE} data. The vertical bars indicate
uncertainties at 1$\sigma$ (68\%) confidence level.
}
\label{0332_0115}
\end{figure*}

In the short set of \textit{RXTE} observations that we have analyzed
(where the source luminosity did not change significantly,
see Sect.\,\ref{sec:obs})
the dynamical range of the pulse amplitude variations reached
a factor of $\sim$$1.5$. For this range we defined six amplitude bins 
and extracted six spectra, one for each bin. The X-ray continuum
was modeled using a powerlaw function with an exponential decay
towards higher energies -- the \textit{XSPEC} \texttt{cutoffpl} 
model: $I(E)\propto E^{-\Gamma}\exp{(-E/E_{\rm fold})}$,
where $E$ is the photon energy, $\Gamma$ -- photon index,
and $E_{\rm fold}$ -- exponential folding energy.
To account for a bump-like feature appearing in the residuals between
10 and 20\,keV, we added a broad Gaussian with the 
centroid energy 
$E_{\rm bump}\sim$11\,keV and a width of $\sigma_{\rm bump}\sim$3\,keV. 

\begin{figure}
\resizebox{\hsize}{!}{\includegraphics{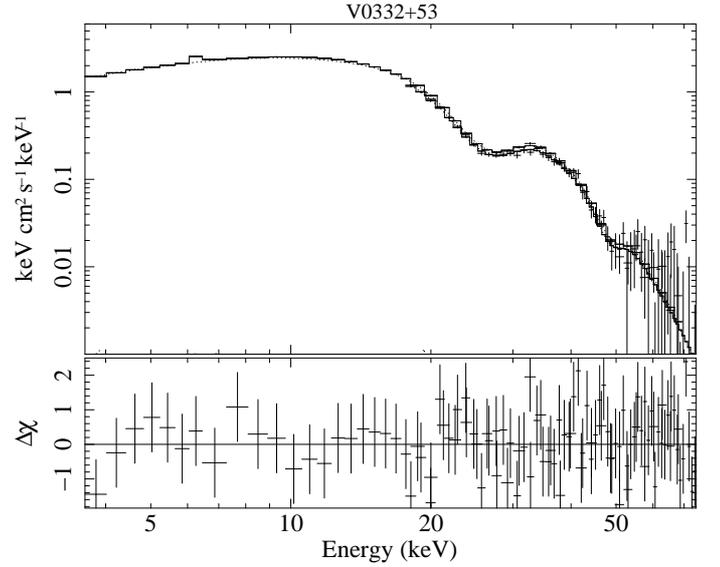}}
\caption{An example of the pulse-amplitude resolved spectrum
of V\,0332+53 obtained with \textit{PCA} and \textit{HEXTE} (from 
the pulse-amplitude bin 3650--3900 cts/s).
The top panel shows the unfolded spectrum modelled with a
cutoff-powerlaw function with two absorption lines (see text).
The bottom panel shows the corresponding residual plot.}
\label{fig:spe}
\end{figure}

The spectra revealed two absorption features,
the fundamental cyclotron line and its first harmonic, which 
we modeled using Gaussian absorption lines as stated above 
(the \emph{XSPEC} \texttt{gabs} multiplicative model: 
$M(E)\propto
\exp{[-\tau_{\rm cyc}/(\sqrt{2\pi}\sigma_{\rm cyc})
\cdot\exp(-0.5(E-E_{\rm cyc})^2/\sigma_{\rm cyc}^2)]}$,
where $E_{\rm cyc}$, $\sigma_{\rm cyc}$, and $\tau_{\rm cyc}$ are 
the centroid energy, width and optical depth of the line,
respectively). 
To account for uncertainties in the absolute normalization
of the flux in different instruments, we introduced in our model
a multiplicative factor {\boldmath $F$} which was fixed to 1.0 for \textit{PCA}
and left free for \textit{HEXTE}.
In all pulse-amplitude bins the best-fit value of $F$ for \textit{HEXTE}
was around $0.90(1)$ indicating a 10\% difference in the absolute flux 
normalization between the two instruments. Similar differences 
(from 10 to 14\%) were found for all other \textit{RXTE} observations
analyzed in this work. 
The energy range of the spectra was set to 3.5--60\,keV for \textit{PCA}
(see Sect.\,\ref{sec:technique} about the usage of \textit{PCA} at
high energies) and to 17--75\,keV for \textit{HEXTE}.
The best-fit spectral parameters for each pulse-phase amplitude bin
together with the 1$\sigma$-uncertainties and the corresponding
reduced $\chi^2$ values are summarized in Table\,\ref{tab:v0332}.
The parameters of the first cyclotron line harmonic are denoted
with ``1'' in the subscript.
Figure\,\ref{fig:spe} shows one of our pulse-amplitude resolved
spectra of V\,0332+53 modeled with the described function.

\begin{table*}
\caption{Best fit spectral parameters of V\,0332+53 for different
pulse-amplitude bins. The indicated uncertainties are at 
1$\sigma$ (68\%) confidence level.}
\label{tab:v0332}
\centering
\renewcommand{\arraystretch}{1.3}
\begin{tabular}{l | c c c c c c}
\hline\hline
Pulse amplitude & & & & & & \\
bin [PCA cts/s] & 2800--3400& 3400--3650& 3650--3900& 3900--4100& 4100--4300& 4300--5100\\
\hline
& & & & & & \\
$\Gamma$ & $-0.621_{-0.010}^{+0.009}$ & $-0.609_{-0.008}^{+0.009}$ & $-0.593_{-0.007}^{+0.008}$ & $-0.586_{-0.008}^{+0.008}$ & $-0.600_{-0.011}^{+0.011}$ & $-0.591_{-0.013}^{+0.014}$\\
$E_{\rm fold}$\,[keV] & 5.79 (fixed) & 5.79 (fixed) & 5.79 (fixed) & 5.79 (fixed) & 5.79 (fixed) & 5.79 (fixed)\\
$E_{\rm bump}$\,[keV] & $11.64_{-0.26}^{+0.29}$ & $11.39_{-0.23}^{+0.25}$ & $11.12_{-0.23}^{+0.26}$ & $10.97_{-0.24}^{+0.27}$ & $10.99_{-0.39}^{+0.41}$ & $10.34_{-0.55}^{+0.51}$\\
$\sigma_{\rm bump}$\,[keV] & 2.65 (fixed) & 2.65 (fixed) & 2.65 (fixed) & 2.65 (fixed) & 2.65 (fixed) & 2.65 (fixed)\\
$E_{\rm Fe}$\,[keV] & 6.4 (fixed) & 6.4 (fixed) & 6.4 (fixed) & 6.4 (fixed) & 6.4 (fixed) & 6.4 (fixed)\\
$\sigma_{\rm Fe}\,[keV]$ & 0.01 (fixed) & 0.01 (fixed) & 0.01 (fixed) & 0.01 (fixed) & 0.01 (fixed) & 0.01 (fixed)\\
$E_{\rm cyc}$\,[keV] & $26.27_{-0.05}^{+0.05}$ & $26.21_{-0.04}^{+0.04}$ & $26.05_{-0.04}^{+0.04}$ & $26.01_{-0.05}^{+0.05}$ & $25.98_{-0.06}^{+0.06}$ & $25.85_{-0.08}^{+0.08}$\\
$\sigma_{\rm cyc}$\,[keV] & $3.65_{-0.07}^{+0.08}$ & $3.61_{-0.07}^{+0.07}$ & $3.64_{-0.06}^{+0.06}$ & $3.74_{-0.07}^{+0.07}$ & $3.97_{-0.10}^{+0.10}$ & $4.04_{-0.12}^{+0.13}$\\
$\tau_{\rm cyc}$ & $11.54_{-0.31}^{+0.32}$ & $11.11_{-0.27}^{+0.28}$ & $10.83_{-0.24}^{+0.25}$ & $10.84_{-0.28}^{+0.29}$ & $11.41_{-0.38}^{+0.40}$ & $11.23_{-0.47}^{+0.50}$\\
$E_{\rm cyc,1}$\,[keV] & $48.96_{-0.92}^{+0.95}$ & $47.71_{-0.62}^{+0.65}$ & $48.56_{-0.68}^{+0.71}$ & $48.98_{-1.02}^{+1.08}$ & $48.15_{-0.80}^{+0.83}$ & $47.76_{-0.90}^{+0.97}$\\
$\sigma_{\rm cyc,1}$\,[keV] & 3 (fixed) & 3 (fixed) & 3 (fixed) & 3 (fixed) & 3 (fixed) & 3 (fixed)\\
$\tau_{\rm cyc,1}$ & $4.7_{-1.2}^{+1.6}$ & $4.9_{-0.9}^{+1.0}$ & $4.3_{-0.8}^{+1.0}$ & $3.3_{-0.9}^{+1.1}$ & $5.1_{-1.1}^{+1.4}$ & $6.0_{-1.5}^{+1.9}$\\
$\chi^2_{\rm red}$/d.o.f. & 1.2/96 & 1.5/96 & 1.5/96 & 1.3/96 & 0.8/96 & 1.5/96\\
\hline
\end{tabular}
\end{table*}

Some spectral parameters which did not show any significant variation
with pulse-amplitude 
were fixed to their averaged (over pulse-amplitude bins) values, 
as indicated in Table\,\ref{tab:v0332}.
The photon index $\Gamma$ and the exponential folding energy $E_{\rm fold}$
were found to be strongly coupled (showing
intrinsic positive correlation with each other), 
i.e. the quality of the data did
not allow us to determine their variation independently. 
$E_{\rm fold}$ showed weaker variability with pulse
amplitude compared to $\Gamma$ and, therefore, was
fixed to its average value. Thus, we only explored the variation
of $\Gamma$ as a function of the pulse amplitude.

The data revealed significant changes of the photon index and 
the energy of the fundamental cyclotron line $E_{\rm cyc}$ with
pulse amplitude as can be seen in Fig.\,\ref{0332_0115} (left column).
The horizontal axis of the plots represents the
pulse amplitude (determined in the way described in Sect.\,\ref{sec:techn})
measured in \textit{PCA} count rate summed over all energy channels 
(2--80\,keV, see above)
and normalized to one \textit{PCU}. The horizontal error-bars indicate
the width of the corresponding pulse amplitude bins. An almost linear
decrease of $E_{\rm cyc}$ with pulse height is clearly seen as well
as a somewhat less significant softening of the spectrum (reflected
by an increase of the photon index $\Gamma$).
The dotted lines represent linear fits to the data. The measured slope of the
$E_{\rm cyc}$ -- pulse-height dependence is $(-2.84\pm
0.50)\times 10^{-4}$\,keV/(cts/s). 
The Pearson linear correlation coefficient is $-0.98$ with the
corresponding two-sided null-hypothesis probability
(the probability to find the correlation by chance in a non-correlated
data sample) $\sim$4$\times10^{-4}$ 
indicating a highly significant anti-correlation.
The slope of the $\Gamma$ -- pulse-height
dependence is $(2.21\pm 0.88)\times 10^{-5}$\,(cts/s)$^{-1}$.
The Pearson linear correlation coefficient with the associated
two-sided null-hypothesis probability are in this case $\sim$$0.77$ and
$\sim$$0.07$, respectively. The (positive) correlation in this case is not very
significant, but still might indicate a possible physical relation
between the parameters.

\begin{figure*}
\centering
\includegraphics[width=17cm]{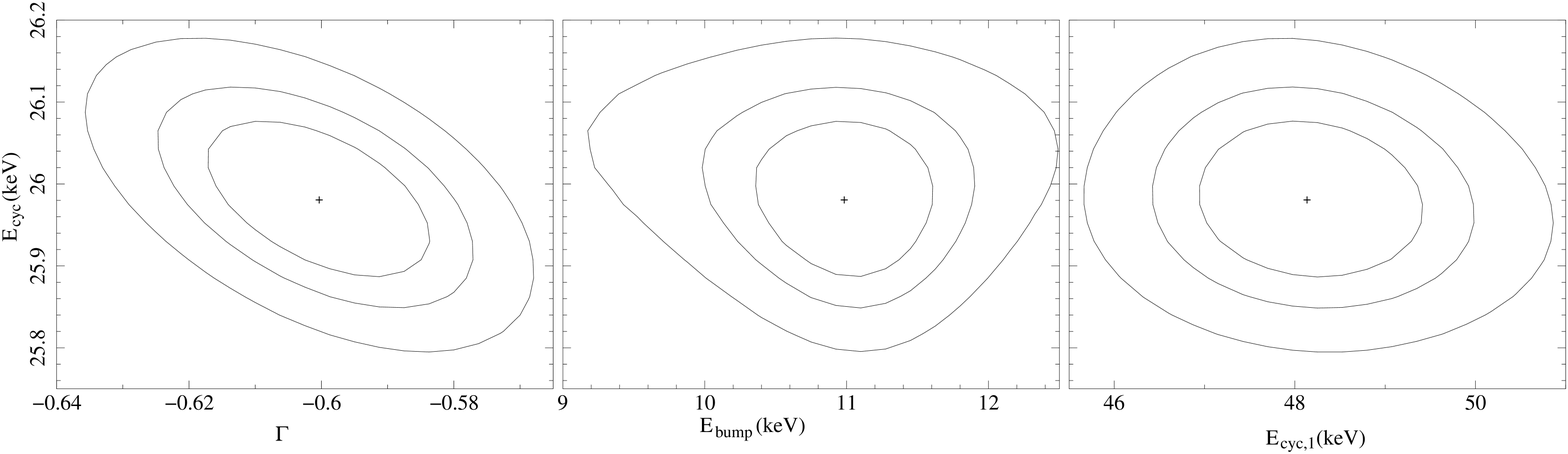}
\caption{$\chi^2$-contour plots for some pairs of spectral parameters
of V\,0332+53. The $\chi^2$-minima are well
defined and the parameters show no degeneracy.
To generate the plots the spectrum from the pulse-amplitude bin
3650--3900 cts/s was used.}
\label{fig:con}
\end{figure*}

To check the reliability of the shown dependencies, we performed spectral 
fits with alternative spectral functions for the continuum and cyclotron
features: \texttt{powerlaw}$\times$\texttt{highecut} 
($I(E)\propto E^{-\Gamma}$ below $E_{\rm cutoff}$ and 
$E^{-\Gamma}\exp{[-(E-E_{\rm cutoff})/E_{\rm fold}]}$ above $E_{\rm cutoff}$)
instead of
\texttt{cutoffpl} and \texttt{cyclabs} (Lorenzian absorption line) instead
of \texttt{gabs}.
For all combinations of the continuum and line models the variations
of $\Gamma$ and $E_{\rm cyc}$ with pulse amplitude emerge with similar
significance.

As mentioned above, we checked our fits for any possible statistical
(model-dependent) correlations of parameters using $\chi^2$-contour plots for
different parameter pairs. All free fit parameters were found
to be well decoupled (i.e. showing no or only a weak statistical
dependence). In all cases the $\chi^2$ minimum and the corresponding
confidence range are well defined.
Examples of the $\chi^2$-contour plots for some parameters
are shown in Fig.\,\ref{fig:con}.

The averaged pulse profile of the source shows two almost equally strong
peaks (Fig.\,\ref{pp}). As described in Sect.\,\ref{sec:technique},
for our study we selected a phase interval including the first 
(the highest) peak. However, we repeated our analysis also for the second
peak and found very similar dependencies of the spectral parameters
on pulse amplitude.

The revealed variation of $E_{\rm cyc}$ agrees with the negative 
correlation of the line energy with the \emph{averaged} (over many pulse
cycles) flux level
during the rise and decay of the strong outbursts reported for this source
(see above). We stress, however, that in our analysis the correlation
with pulse height appears in a short (less than one day) set of pointings
where the average flux stays roughly constant. The positive correlation of
$\Gamma$ with pulse height, if it is real, is also consistent with the gradual
decrease of the averaged photon index observed 
during the decay of the outburst which can be derived from
Table\,2 of \citet{Mowlavi:etal:06}.

\subsection{4U\,0115+63}

Another member of the BeXRB class, 4U\,0115+63 has a pulsation period
of $\sim$3.6\,s \citep{Cominsky:etal:78}. 
Up to five cyclotron line harmonics have been detected in the X-ray
spectrum of the source with the fundamental line at $\sim$11-16\,keV
having the lowest energy among accreting pulsars
\citep[][and references therein]{Heindl:etal:99,
Santangelo:etal:99,Ferrigno:etal:09}.
Like in V\,0332+53, the fundamental line in 4U\,0115+63 was found 
to increase during the decay phase of the giant outbursts 
\citep{Mihara:etal:04,Tsygankov:etal:07}.
For our analysis we used the \textit{RXTE} observations of the intense
giant outburst of the source in 1999 when the bolometric luminosity
exceeded $10^{38}$\,erg\,s$^{-1}$ assuming a distance of 7\,kpc
\citep{Tsygankov:etal:07}. The data cover roughly 0.5 day of the
outburst close to its maximum.
In our pulse-amplitude resolved spectra we detected (and modeled 
using the \texttt{gabs} \textit{XSPEC} model, see above) the fundamental
line and its three harmonics. Following \citet{Tsygankov:etal:07} and
\citet{Ferrigno:etal:09}, we used the
\texttt{powerlaw}$\times$\texttt{highecut} model for the spectral
continuum. 

\begin{figure*}
\centering
\includegraphics[width=17cm]{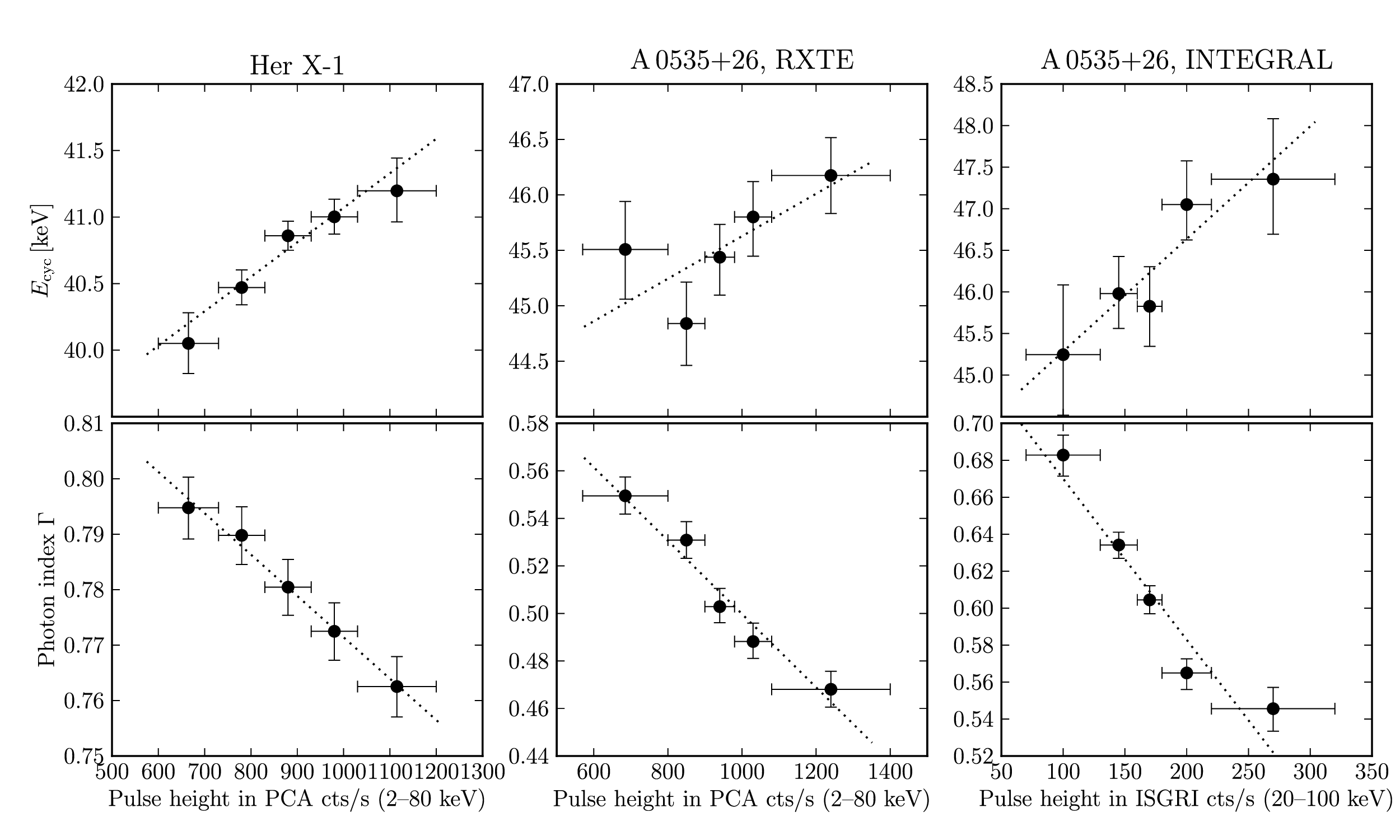}
\caption{The variation of the fundamental cyclotron line centroid energy 
$E_{\rm cyc}$ (top) and the photon index $\Gamma$ (bottom)
with pulse amplitude measured in Her X-1 using the 
\textsl{RXTE} (left column)
and in A\,0535+26 using \textsl{RXTE} (middle column), and using 
\textit{INTEGRAL} (right column). The vertical bars indicate
uncertainties at 1$\sigma$ (68\%) confidence level.
The dotted lines represent linear fits to the data points.
}
\label{her_0535}
\end{figure*}

The amplitude of variations in the pulse height exhibited by 4U\,0115+63
in our data sample reached a factor of $\sim$$2.5$. Like in the case
V\,0332+53 the data quality allowed us to divide the pulse-height 
distribution into six bins and explore the variations in
the centroid energy of the fundamental cyclotron line $E_{\rm cyc}$
and the photon index $\Gamma$ with pulse height 
(Fig.\,\ref{0332_0115}, right column). 
The exponential folding 
energy $E_{\rm fold}$ was again found to be coupled with $\Gamma$. As 
in the case of V\,0332+53 we fixed $E_{\rm fold}$ to its mean 
(over pulse-height bins) value, 
8.2\,keV, so that the 
softening/hardening of the spectral
continuum is described by the changes in $\Gamma$
and the cutoff energy $E_{\rm cutoff}$ (the latter, however, did not
show any significant dependence on pulse amplitude). 
Parameters of the cyclotron line harmonics showed no 
significant variation with pulse amplitude.

As can be seen in the top right plot of Fig.\,\ref{0332_0115}, 
the energy of the fundamental
cyclotron line decreases with increasing pulse amplitude
(probably showing some saturation at highest amplitudes, see below),
similarly to V\,0332+53.
The source also shows softening of the spectrum
(increase of the photon index)
with increasing pulse height (bottom right plot in 
Fig.\,\ref{0332_0115}), which is also similar to V\,0332+53.

As before, the dotted lines represent linear fits to the data points. The measured slopes
are $(-1.32\pm 0.38)\times 10^{-4}$\,keV/(cts/s)
for the $E_{\rm cyc}$ -- pulse-height dependence and 
$(4.27\pm 0.68)\times 10^{-5}$\,(cts/s)$^{-1}$
for the $\Gamma$ -- pulse-height dependence, indicating a significant
variation. The Pearson correlation coefficient and the associated
two-sided null-hypothesis probability are $\sim$$-0.58$ and
$\sim$$0.22$ respectively for $E_{\rm cyc}$, and $\sim$$0.70$ and
$\sim$$0.12$ for $\Gamma$. As can be seen, the linear correlation inspection
indicate rather sparse correlations of the parameters. We note,
however, that the formal linear correlation analysis does not take
into account the uncertainties of the parameters (which are known in
our case) and is not very reliable in case of a limited sample
(a few points). The formal linear fits taking into account the
uncertainties, nevertheless, indicate significant variations (see
the slope values above).  The two rightmost points in the 
$E_{\rm cyc}$ -- pulse-height plot (Fig.\,\ref{0332_0115}, top right)
might indicate some saturation of the downward trend in the cyclotron
line energy. However, as the uncertainties associated with the points are
larger compared to the other points, the  apparent "flattening" might
result from poor statistics. The leftmost data point on the 
$\Gamma$ -- pulse-height plot (Fig.\,\ref{0332_0115}, bottom right)
significantly deviates from a linear upward trend formed by the other
data points. Considering the relatively small uncertainties of the
data values, this might indicate a more complicated (than the linear)
relation between $\Gamma$ and the pulse amplitude in this source.

Like in V\,0332+53, the decrease of the line energy with the pulse amplitude
(which we see on the time scale of single 
X-ray pulsations) is in line with the negative correlation
of $E_{\rm cyc}$ with the average flux measured 
on a much longer time scale during the decay of the source's
outbursts. 

For 4U\,0115+63 we did not find any alternative spectral function
that would provide an acceptable fit of the spectral continuum. However,
we checked our spectral fits using Lorentzian absorption lines to model
CRSFs instead of the Gaussian lines. The reported
effects (variations of $E_{\rm cyc}$ and $\Gamma$) appear equally 
significant in the spectral fits with both types of absorption lines. 

\subsection{Her X-1}

Unlike the V\,0332+53 and 4U\,0115+63, Her~X-1 is a persistent X-ray source. 
As stated in Sect.\,\ref{sec:obs}, the observed regular changes of the source
flux are related to the periodic obscuration by matter in the accretion
flow, whereas the intrinsic luminosity of the source 
(estimated to be $\sim$2$\times 10^{37}$\,erg\,s$^{-1}$ for a distance
of 7\,kpc, \citealt{Reynolds:etal:97})
remains roughly constant
(possibly exhibiting, however, long-term changes, see below).
The pulsation period of Her~X-1 is $\sim$1.24\,s, which is the shortest
period in our sample. Being historically the first 
source where a cyclotron feature was detected \citep{Truemper:etal:78},
Her~X-1 has the most regular and continuous records of the
line energy among all accreting neutron stars \citep[e.g.][]{Staubert:etal:07,
Klochkov:etal:08}. Analyzing the \textit{RXTE} observations of the source
spread over $\sim$10 years, \citet{Staubert:etal:07} found that
the cyclotron line centroid energy is \emph{positively} correlated
with the X-ray luminosity which apparently slowly varies with
the amplitude of a factor of two between different main-on states
(the authors used the maximum main-on flux as a measure of the source
luminosity).
Such a behavior is opposite to that observed in the outbursts
of the transient sources V\,0332+53 and 4U\,0115+63 (see above)
making Her~X-1 particularly interesting for our 
pulse-amplitude-resolved
study.

We found that the amplitude of the single pulses during the selected observation
(covering a large fraction of a main-on, see Sect.\ref{sec:obs})
varies by a factor of $\sim$2. For spectral extraction, we selected five
pulse-amplitude bins. 
The spectrum was modeled with the function
used by \citet{Staubert:etal:07} (and in most previous
spectral studies of Her~X-1) --
\texttt{powerlaw$\times$highecut} with a Gaussian absorption line 
to model the CRSF. The exponential cutoff energy $E_{\rm cutoff}$
was found to be strongly coupled with the photon index 
$\Gamma$ and, therefore, was fixed to its average (over all pulse-amplitude
bins) value, 21.4\,keV. 
As a result of our analysis, we were able to detect
a strong \emph{positive correlation} between $E_{\rm cyc}$ and the pulse
amplitude during a \emph{single} main-on state (Fig.\,\ref{her_0535}, top left
panel). 
The slope of the best-fit linear relation (dotted line) is
$(2.59\pm 0.58)\times 10^{-3}$\,keV/(cts/s).
The Pearson linear correlation coefficient is $\sim$$0.97$
corresponding to the probability to find the correlation by chance (two-sided)
of only $\sim$6$\times 10^{-3}$ -- highly significant positive correlation.
This confirms the dependence found by
\citet{Staubert:etal:07} using a completely different approach.
Additionally, we detected a \emph{negative} correlation between the photon
index $\Gamma$ and the pulse amplitude 
(Fig.\,\ref{her_0535}, bottom left panel) 
which is contrary to V0332+53 and 4U\,0115+63 
(see Fig.\,\ref{her_0535} and the previous subsections).
The slopes of the best-fit linear relation (dotted line)
between $\Gamma$ and the pulse height is
$(-7.46\pm 1.56)\times 10^{-5}$\,(cts/s)$^{-1}$.
The Pearson linear correlation coefficient and the corresponding
two-sided null-hypothesis probability are
$\sim$$-0.99$ and $\sim$$4\times 10^{-4}$ respectively -- again, a highly
significant (negative) correlation.

As in the case of 4U\,0115+63, the \texttt{powerlaw$\times$highecut} function
was the only suitable continuum model. Thus, we checked stability of our
results only relative to the choice of the cyclotron model: 
\texttt{gabs} vs. \texttt{cyclabs}. The found correlations are perfectly
seen with both models.

\subsection{A\,0535+26 \label{sec:a0535}}

A\,0535+26 is another BeXRB that sporadically shows intense outbursts. 
It has, however, a much longer pulsation period
compared to V\,0332+53 and 4U\,0115+63 -- around 103\,s
\citep{Finger:etal:94}. Two absorption features, interpreted
as cyclotron lines (fundamental and first harmonic), have been observed 
in the source X-ray spectrum at $\sim$45\,keV and $\sim$100\,keV
\citep[][and references therein]{Caballero:etal:08}.
Contrary to the sources described above, no clear variations of the cyclotron 
line energy with flux have so far been reported \citep[][]{Caballero:etal:08}.

For A\,0535+26 we used the data obtained simultaneously
with \textsl{RXTE} and \textsl{INTEGRAL} covering $\sim$1 day around the
maximum of the normal outburst in 2005 when the source luminosity 
reached roughly $\sim$10$^{38}$\,erg\,s$^{-1}$ \citep{Caballero:etal:07}.
Following \citet{Caballero:etal:07,Caballero:etal:08}, we used
the \textit{XSPEC} \texttt{cutoffpl} model 
(see Sect.\,\ref{subsec:v0332}) to describe the spectral
continuum and a Gaussian absorption line to model the fundamental CRSF
(the first harmonic was not detected in our pulse-amplitude resolved
spectra). 
In case of \textit{INTEGRAL} we set the energy range to 4--30\,keV
for \textit{JEM-X} and to 18--90\,keV for \textit{IBIS/ISGRI}.
As for \textit{RXTE}, 
we introduced in our spectral model
a multiplicative factor {\boldmath $F$} which was fixed to 1.0 for 
\textit{ISGRI}
and left free for \textit{JEM-X} to account for
uncertainties in the absolute flux normalization in the two instruments.
In all pulse-amplitude bins the best-fit value of $F$ for \textit{ISGRI}
was around $0.96(2)$ indicating a 4\% difference in the absolute flux 
normalization between the two instruments. 
As for the other sources, due to the strong intrinsic coupling
between the photon index $\Gamma$ and the exponential folding energy
$E_{\rm fold}$ (both in the \textit{RXTE} and \textit{INTEGRAL} data),
we fixed the latter to its mean value 
(15.8\,keV for \textit{RXTE} and 16.3\,keV for \textit{INTEGRAL})
and explored
only the variation of $\Gamma$.

As can be seen in Fig.\,\ref{her_0535}
the data from both satellites independently show a strong indication of a
\emph{positive} correlation between the fundamental 
cyclotron line energy and the pulse height. 
The horizontal axis of the plots showing the \textit{INTEGRAL} data
represents the pulse amplitude 
measured in units of \textit{IBIS/ISGRI} count rate in the
20--100\,keV range. 
The slopes of linear fits to the data (as before, indicated with the
dotted lines) are $(1.92\pm 0.92)\times 10^{-3}$\,keV/(PCA cts/s)
for the \textsl{RXTE} observations (Fig.\,\ref{her_0535}, top middle)
and $(1.36\pm 0.53)\times 10^{-2}$\,keV/(ISGRI cts/s)
for the \textsl{INTEGRAL} observations (Fig.\,\ref{her_0535}, top right).
The Pearson correlation coefficients with the associated two-sided
null-hypotheses probability values are
$\sim$$0.70$ and $\sim$$0.19$ respectively for the \textsl{RXTE}
data (rather sparse correlation) and $\sim$$0.94$ and $\sim$$0.02$
for the \textsl{INTEGRAL} data. We repeat, however, that the similar
relations are found by the two satellites independently, 
supporting the correlation.
Our finding is particularly interesting as no corresponding ``long-time-scale
correlation'' of $E_{\rm cyc}$ with flux has been reported (see above). 
Our results make A\,0535+26 only the second source (after Her~X-1) 
showing positive correlation of the cyclotron line energy with flux.

Additionally, the data from both satellites show hardening of the spectrum
(decrease in $\Gamma$) with increasing pulse height, as shown in
the bottom panels of the middle and the right column in 
Fig.\,\ref{her_0535}. 
The slopes of linear fits to the data 
(dotted lines) are $(-1.54\pm 0.19)\times 10^{-4}$\,keV/(PCA cts/s)
for the \textsl{RXTE} observations (Fig.\,\ref{her_0535}, bottom middle)
and $(-8.71\pm 0.86)\times 10^{-2}$\,keV/(ISGRI cts/s)
for the \textsl{INTEGRAL} observations (Fig.\,\ref{her_0535}, bottom right).
The Pearson correlation coefficients with the associated two-sided
null-hypotheses probability values are
$\sim$$-0.98$ and $\sim$$4\times 10^{-3}$ respectively for the \textsl{RXTE}
data and $\sim$$-0.96$ and $\sim$$0.01$
for the \textsl{INTEGRAL} data -- significant correlation in both cases.
This (anti-)correlation
is contrary to the one observed in V\,0332+53 and 4U\,0115+63 but similar to
the one found in Her~X-1, which also shows a
positive correlation between $E_{\rm cyc}$ and pulse amplitude.
The negative correlation of $\Gamma$ with pulse height in A\,0535+26
is also in line with Fig.\,4 of \citet{Caballero:etal:08} showing
larger values of the photon index at lower luminosities (at the 
beginning and the end of the outburst).

We note, that it is not possible to compare directly the
variations  of the spectral parameters 
measured with \textit{INTEGRAL} and \textit{RXTE} 
as a function of pulse amplitude with each other.
The high-resolution light curves used for the pulse selection were obtained
with \textit{INTEGRAL/IBIS/ISGRI} and \textit{RXTE/PCA} 
in different energy bands,
(\textit{PCA} is mostly sensitive below $\sim$35\,keV, while \textit{IBIS/ISGRI}
-- above 20\,keV, see Sect.\,\ref{sec:obs}), which leads to 
systematically different pulse-height distributions. 
Additionally, as it was mentioned in Sect.\,\ref{sec:technique},
insufficient intercalibration accuracy of the two observatories
might
lead to systematic uncertainties in the absolute
values of $E_{\rm cyc}$ and $\Gamma$ measured with the two instruments.
Nevertheless, the \emph{relative}
variations of the spectral parameters with pulse amplitude can still be 
determined with the two satellites regardless of cross-calibration problems. 
Thus, the similarity
of the parameters behavior found independently 
with \textit{RXTE} and \textit{INTEGRAL} indicates
that the obtained effects indicate real physics.

As with other sources, we checked the found correlations using 
the Lorentzian line profile instead of the Gaussian one and found them 
to be independent with respect to the choice of the line model. 

\section{Discussion}
\label{sec:discussion}

The revealed dependencies of the spectral parameters on pulse
amplitude in our sample of bright X-ray pulsars are mostly consistent 
with the long-term spectral changes related to the variability
of the averaged 
luminosity of the sources (e.g. during the outbursts) reported previously,
i.e. the correlations between the spectral parameters and flux
have the same sign.
We note, however, that it is generally impossible to compare directly
our correlations with those found on the basis of the long-term flux variations.
Our pulse amplitude reflects the flux in a narrow pulse-phase interval, 
thus its value and variability range substantially differ from those of 
the pulse-averaged flux. Therefore, the individual pulse heights 
cannot easily be converted to the source luminosity as 
it was done for the pulse-averaged flux in the previous works. 
The spectral
parameters in our analysis were also extracted from the narrow pulse-phase 
interval and are, therefore, different from the pulse-averaged values.
The variability of the average (over many pulse cycles) pulse profile
along the outbursts in the previously 
reported ``long-term'' analyses would further
complicate the comparison. 

It has been argued by several authors that
the long-term spectral variability of the cyclotron line energy
reflects the changes in the accretion column structure,
namely the variable height of the X-ray emitting region above
the neutron star surface as a function of the mass accretion rate
$\dot M$ \citep{Mihara:etal:98,Mowlavi:etal:06,Staubert:etal:07}.
As the variability revealed in our work apparently takes place on the
time scale of single X-ray pulsations, we have to conclude that
the characteristic time scale of the changes in the emitting
structure (in response to variable $\dot M$ reflected by the pulse height) 
are of the order of a few seconds or shorter. Any X-ray spectra 
accumulated over many subsequent pulsation cycles of a source, 
therefore, provide only an average characterization
of the accretion structure over a range of its states corresponding to
different local accretion rates.

\begin{table*}
\caption{Correlations of spectral parameters with pulse height}
\label{cor}
\centering          
\begin{tabular}{l | c c c | c c c}     
\hline\hline   
Source name & \multicolumn{3}{c|}{$E_{\rm cyc}$ vs. pulse height} & 
\multicolumn{3}{c}{$\Gamma$ vs. pulse height}\\
\cline{2-7} 
& slope  & Pearson  & 2-sided & slope & Pearson & 2-sided \\
&  [keV/(cts/s)]  & cor. coeff. & $P$-value & [(cts/s)$^{-1}$]  & cor. coeff. & $P$-value \\
\hline        
& & & & & & \\
    V\,0332+53  & $(-2.84\pm 0.50)\times 10^{-4}$ & $-0.98$ & $4\times10^{-4}$ & 
                             $(2.21\pm 0.88)\times 10^{-5}$   & $+0.77$ & $0.07$ \\
    4U\,0115+63 & $(-1.32\pm 0.38)\times 10^{-4}$& $-0.58$ & $0.22$ &
                             $(4.27\pm 0.68)\times 10^{-5}$   & $+0.70$ & $0.12$ \\
    Her X-1         & $(2.59\pm 0.58)\times 10^{-3}$  &  $+0.97$& $6\times 10^{-3}$ &
                             $(-7.46\pm 1.56)\times 10^{-5}$& $-0.99$ & $4\times 10^{-4}$  \\
A\,0535+26, XTE &$(1.92\pm 0.92)\times 10^{-3}$& $+0.70$& $0.19$ &
                              $(-1.54\pm 0.19)\times 10^{-4}$& $-0.98$& $4\times 10^{-3}$ \\
A\,0535+26, INT. &$(1.36\pm 0.53)\times 10^{-2}$& $+0.94$ & $0.02$ &
                               $(-8.71\pm 0.86)\times 10^{-2}$& $-0.96$& $0.01$ \\
\hline
\end{tabular}
\end{table*}

\subsection{Two regimes of accretion}
\label{sec:regimes}

Two of the four accreting pulsars studied in this work exhibit a negative
correlation of the fundamental cyclotron line energy and a positive 
correlation of the photon index with pulse amplitude. In the other
two sources the correlations are in the opposite direction: a positive
correlation of $E_{\rm cyc}$ and a negative
correlation of $\Gamma$ with pulse height. 
The correlations quantified by the values of the linear slope and the
linear correlation coefficients (see the previous section) are
summarized in Table\,\ref{cor}.
The two groups of pulsars apparently have different physical
conditions inside their X-ray emitting structure leading to
the different behavior of the spectral parameters.


The ``long-term'' negative correlation of the cyclotron line energy
with the X-ray flux during the rises and decays of the outbursts of
V\,0332+53 and 4U\,0115+63 has been attributed to the variable height of
the radiative shock above the accretor surface where a substantial part of the 
kinetic energy possessed by inflowing matter transforms into radiation
\citep{Mihara:etal:98,Mihara:etal:04,Mowlavi:etal:06,Tsygankov:etal:07}.
The height of the shock in these pulsars is believed to be directly 
proportional to the local mass accretion rate $\dot M$ 
\citep{Basko:Sunyaev:76}. We argue that the negative correlation
of $E_{\rm cyc}$ with pulse amplitude reflects the same physical effect
(changing of the shock height with $\dot M$) but on much shorter
time scales. Indeed, the pulse height most probably reflects
the local $\dot M$ which is highly variable on short time scales
leading to the observed pulse-to-pulse variability.  On the other hand,
the ``reaction time'' of the emitting structure determined by
the characteristic time scales at which the kinetic energy of
infalling matter is converted into radiation and the latter diffuses through
the column has been shown to be short enough -- less than $10^{-6}-10^{-7}$\,s 
\citep{Orlandini:Boldt:93,Morfill:etal:84}. The softening of the 
spectral index with pulse amplitude can also be qualitatively 
understood in this framework.
According to the basic model of the accretion column in 
\citet{Basko:Sunyaev:76} the plasma temperature in the column monotonically
decreases with height (up to the the radiative shock). Thus, at higher 
$\dot M$ (i.e. at larger pulse amplitudes) one would expect
more soft photons produced by the lateral walls of a taller column
which would naturally lead to a softer X-ray spectrum, as it is observed.

Generally, the negative $E_{\rm cyc}-\dot M$ correlation is only expected
if the mass accretion rate is high enough for the radiative shock
to form, i.e. if it is above some critical luminosity $L_c$, roughly
corresponding to the local Eddington limit at the polar caps
\citep{Basko:Sunyaev:76,Nelson:etal:93}. If the luminosity is
below $L_c$, a different accretion regime is expected. So far, only
one pulsar, Her~X-1, has been reported to have a statistically significant
positive
$E_{\rm cyc}-\dot M$ correlation \citep{Staubert:etal:07}. The authors
argued that such a correlation is indeed expected at the luminosities
below $L_c$, when infalling matter is stopped by the Coulomb drag and 
collective plasma effects rather than in a radiative shock.
Our pulse-amplitude-resolved analysis
showed that the positive  $E_{\rm cyc}-\dot M$
correlation found by \citet{Staubert:etal:07} on the time scale of
months to years is also present on the time scale of individual
pulses and, thus, supports the idea that in Her~X-1 accretion 
proceeds in the local sub-Eddington regime (contrary to
V\,0332+53 and 4U\,0115+63).

A positive correlation of $E_{\rm cyc}$ with pulse height in A\,0535+26 
which we found here for the first time apparently reveals the second
(after Her~X-1) case of a positive $E_{\rm cyc}-\dot M$ correlation in
an accreting pulsar. A similar to Her~X-1 dependence of $\Gamma$
on pulse amplitude (see Table\,\ref{cor}) additionally supports 
the idea that a different (compared to V\,0332+53 and 4U\,0115+63) 
accretion regime is realized in the two sources. We note here, that the value
of the critical luminosity $L_c$ is expected to vary from one pulsar
to another as it generally depends on the $B$-field strength at the 
neutron star surface and the area of the polar cap (which might depend
on the local magnetic field configuration, i.e. presence of 
higher multipole components etc.). Thus, it is well possible that the
pulsars at a similar luminosity level (as e.g. A\,0535+26 and 4U\,0115+63)
exhibit different accretion regimes.

We note that the small number of sources in our sample 
does not allow us to extend the obtained results (existence
of the two accretion modes described above) to the entire population
of accreting pulsars. Further observations of known cyclotron line
sources together with discoveries of new X-ray pulsars showing
cyclotron line variations would allow one to test our finding on a 
larger sample of pulsars.

\section{Summary and conclusions}
\label{sec:summary}

In the presented work we studied the 
pulse-amplitude-resolved spectral
variability in a sample of four bright accreting pulsars using the 
high-resolution X-ray data taken with \textit{RXTE} and \textit{INTEGRAL}.
Our analysis revealed for the first time the spectral differences
between X-ray pulses of different amplitudes, both in the spectral continuum
and in the resonant cyclotron feature. 

For the pulsars where a negative correlation of the cyclotron line energy
with flux on longer time scales has been reported previously we found a similar 
(negative) correlation on the time scale of singe pulse cycles. 
In Her~X-1 we found a positive correlation of the line energy with
pulse height which is consistent with the long-term correlation
of the CRSF energy and the maximum main-on flux reported by 
\citet{Staubert:etal:07}. Our analysis revealed
a positive correlation of the cyclotron line energy with pulse amplitude
in A\,0535+26 where no correlation of the CRSF energy with flux
has so far been reported.

The pulsars in our sample show two different types
of spectral dependencies on pulse amplitude. We argue that the different
behaviors reflect two distinct accretion regimes (local sub- and 
super-Eddington) realized in different pulsars, as it was previously
proposed to explain the peculiarity of Her~X-1 compared to higher-luminosity
pulsars.

\begin{acknowledgements}
The work was supported by the Carl-Zeiss-Stiftung and by DLR grant BA5027.
DK thanks J\"orn Wilms, Sergey Tsygankov, and Konstantin Postnov
for useful discussions.

We also thank ISSI (Bern, Switzerland) for its hospitality during the 
meetings of our collaboration in the frame of the International Teams program.
\end{acknowledgements}

\bibliographystyle{aa}
\bibliography{refs}

\end{document}